# Unified Modeling of Complex Real-Time Control Systems

He Hai,  Zhong Yi-fang,  Cai Chi-lan
*National CAD Support Software Engineering Research Center,
Huazhong University of Science & Technology,Wuhan,China*
hehai_2000@yahoo.com.cn

## Abstract

*Complex real-time control system is a software dense and algorithms dense system, which needs modern software engineering techniques to design. UML is an object-oriented industrial standard modeling language, used more and more in real-time domain. This paper first analyses the advantages and problems of using UML for real-time control systems design. Then, it proposes an extension of UML-RT to support time-continuous subsystems modeling. So we can unify modeling of complex real-time control systems on UML-RT platform, from requirement analysis, model design, simulation, until generation code.*

## 1. Introduction

Object-oriented (OO) modeling languages, tools, and methods more and more attract the interest of real-time control system developers. UML is an object-oriented industrial standard modeling language, offering many advantages for modeling complex control systems: (1) UML is a standard language, which can be easily understood by customers, software engineers and control engineers. (2) well-known software engineering principles such as information hiding and reusing of software components are supported by the OO paradigm [1]. (3) There are many commercial UML modeling tools for user to choose. UML-RT is an extension of UML for real-time domain, aiming at modeling event-driven real-time systems [2]. But complex real-time control systems contain both time-discrete and time-continuous blocks and additional software components. Presently, modeling these kinds of systems needs use several tools together, such as UML and Simulink. M. Kühl proposes a universal object-oriented modeling method, which translates Simulink into UML, then moves to an UML based toolsuite for real-time code generation [3]. In this case, lots of objects and classes may be generated, and some information may be lost. L. Bichler presents an interesting method which extends capsules for containing two kinds of ports (data ports and signal ports), and associating each state with an arbitrary number of directed equations [4]. Because UML is a foundational discrete language, so this method doesn't work efficiently. This paper introduces some new stereotypes and shows how to extend UML-RT service library framework for supporting time-continuous dataflow model, so that complex control systems can unify modeling on a UML-RT platform.

## 2. UML-RT extensions for control systems

Complex real-time control systems are hybrid systems of time-discrete and time-continuous, whose behaviors can be described by difference equations and differential equations respectively. In UML-RT, difference equations can be integrated into capsule's actions (e.g. transition, entry, exit state). But to differential equations, this kind of integration is infeasible, because these equations must be continuous computed, and UML-RT has a "run-to-complete" semantic. In this paper, we assign event-driven capsule and time-continuous dataflow to different threads. This method separating algorithms from states, making the architecture of software very sound, is a good design pattern, shown in Figure 1.

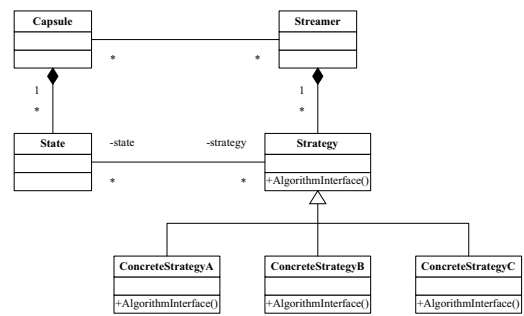

**Figure 1.  Class diagram of state and algorithms**

**Table 1. New stereotypes comparing with UML-RT**

| UML-RT | Extension |
|---|---|
| capsule | streamer |
| port | DPort, SPort |
| connect | flow, relay |
| protocol | flow type |
| state machine, state | slover, strategy |
| Time service | Time |

This paper introduces eight new stereotypes, listed in Table 1. **Streamers** have some same characteristics as capsules. As such, streamers have ports through which they communicate with other objects, and they can contain any number of sub-streamers. Streams are distinguished from capsules by their behaviors, which is implemented by a solver through computing equations. Streamers have two kinds of ports: data ports (**DPorts**) and signal ports (**SPorts**), which denoted by circle and square respectively. Data ports carrying dataflow, have some kind of data type (**flow type**). To connect two DPorts, the output DPorts' flow type must be a subset of the input DPorts flow type. **Relay** is used as a relay point which generates two similar flows from a flow. SPorts convey signal message, which associated with a protocol. Streamers can communicate with capsules through SPorts. In a streamer, there is a **solver** responsible for receiving signal from SPorts and data from DPorts and operating system services, modifying parameters, computing equations, and sending out the results. Timing in UML-RT is unpredictable. In this paper, we introduce a **Time** stereotype, which is a continuous variable, can be used as simulation clock. Figure 2 shows the abstract syntax of streamers.

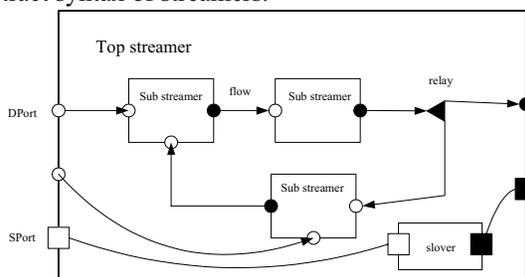

**Figure 2. Abstract syntax of streamers**

To make modeling intuitionistic, some extensions are introduced to capsules, making them also have DPorts and SPorts. But in capsules, DPorts are only used as relay ports. No data will be processed by capsules. According to same principle, this paper assumes capsules can contain streamers, but streamers don't contain any capsule. In the model, we can use any number of streamers, which are assigned to one or several threads during implementation. The structure of the extension shows in figure 3.

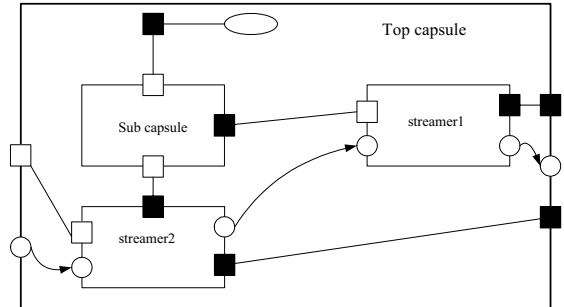

**Figure 3. Structure of extensions**

During implementation, capsules and streamers are assigned to different threads. Communication between capsules and streamers is realized by communication mechanism of threads. Streamer also can use operating system services to get and send data from devices. Behavior of capsules is described by state machine, and behavior of streamers is carried out by solvers through computing the equations.

## 3. Conclusions

This paper proposes an extension for UML-RT services library framework to support time-continuous dataflow modeling. It assigns event-driven part and time-continuous part of systems to different threads, and makes use of operating system communication mechanism as a channel between capsules and streamers. As we can see, this method makes the architecture of complex control system very sound, and easy to realize on existing UML-RT platforms.